\documentstyle[11pt,aaspp4,psfig]{article}

\begin{document}

\title{The Progenitor of the New COMPTEL/ROSAT Supernova Remnant in 
Vela\footnote{To appear in The Astrophysical Journal, Letters, Vol. 514}}

\author{Wan Chen$^{1,2}$ and Neil Gehrels$^2$} \affil{$^1$ Department of
Astronomy, University of Maryland, College Park, MD 20742 \\ $^2$
NASA/Goddard Space Flight Center, Code 661, Greenbelt, MD 20771 \\
chen@milkyway.gsfc.nasa.gov, gehrels@lheavx.gsfc.nasa.gov}

\begin{abstract}

We show that (1) the newly discovered supernova remnant (SNR), GRO
J0852--4642/RX J0852.0--4622, was created by a core-collapse supernova of a
massive star, and (2) the same supernova event which produced the $^{44}$Ti
detected by COMPTEL from this source is probably also responsible for a large
fraction of the observed $^{26}$Al emission in the Vela region detected by
the same instrument.  The first conclusion is based on the fact that the
remnant is currently expanding too slowly given its young age for it to be
caused by a Type Ia supernova.  If the current SNR shell expansion speed is 
greater than 3000 km s$^{-1}$, a $15 M_\odot$ Type II supernova with a 
moderate kinetic energy exploding at about 150 pc away is favored.  If the 
SNR expansion speed is lower than 2000 km s$^{-1}$, as derived naively from 
the X-ray data, a much more energetic supernova is required to have occurred 
at $\sim250$ pc away in a {\it dense} environment at the edge of the Gum 
nebula.  This progenitor has a preferred ejecta mass of $\le10 M_\odot$ and 
therefore, it is probably a Type Ib or Type Ic supernova.  However, the 
required high ambient density of $n_H \ge 100$ cm$^{-3}$ in this scenario is 
difficult to reconcile with the regional CO data.  A combination of our 
estimates of the age/energetics of the new SNR and the almost perfect 
positional coincidence of the new SNR with the centroid of the COMPTEL $
^{26}$Al emission feature of the Vela region strongly favors a causal 
connection.  If confirmed, this will be the first case where both $^{44}$Ti 
and $^{26}$Al are detected from the same young SNR and together they can be 
used to select preferred theoretical core-collapse supernova models.

\end{abstract}


\section{Introduction}

Recently, the COMPTEL instrument aboard the {\it Compton Gamma Ray
Observatory} discovered a new supernova remnant (SNR), GRO J0852--4642, as a
strong source of $^{44}$Ti 1.16 MeV line emission (\cite{rf:iafea98}).  The
new source is located in the Constellation Vela in the direction of $l\sim
266^\circ$ and $b\sim -1^\circ$.  Due to the short lifetime of $^{44}$Ti, 
$\tau = 90.4\pm1.3$ yr (\cite{rf:nebea97}), the fact that COMPTEL can detect
this source at all indicates that it is a young SNR at a distance less than
500 pc with an age probably less than $10\tau \sim 900$ yr.

The COMPTEL $^{44}$Ti source was soon identified with a previously unknown
shell-type SNR in {\it ROSAT} data of the Vela region collected during its
1990 all-sky survey (\cite{rf:ab98}).  Designated as RX J0852.0--4622, the
new SNR is seen within the error circle of the COMPTEL source, but only at
energies $>1.3$ keV.  The shell has a radius $\theta \sim 1^\circ$.  Spectral
analysis reveals that X-rays from the majority of the shell are of thermal
origin with a temperature of $kT = 2.5^{+4.5}_{-0.7}\,\,{\rm keV,}$ which
is significantly greater than that of the surrounding area.  This is why the
shell can be identified only at high energies, and this also indicates a 
young age.  The northern limb of the shell appears brighter and has an even 
higher temperature of $4.7^{+4.5}_{-0.7}$ keV.  This component can also be 
fitted by a power law with photon index $\alpha =-2.6^{+0.3}_{-0.4}$.  In 
addition, the {\it ROSAT} data gives an upper limit of $n_H < 0.04 (d/500$ 
pc)$^{-1/2}$ cm$^{-3}$ to the density of the ambient medium in which the 
progenitor of RX J0852.0-4622 exploded (\cite{rf:ab98}).

We show in this {\it Letter} that, assuming COMPTEL and ROSAT detected the
same source, the above observed properties can place rather stringent
constraints on the nature of the progenitor star of GRO J0852--4642/RX
J0852.0--4622 (hereafter GRO/RX J0852).

The supernova which produced the observed radioactive $^{44}$Ti would also 
have synthesized about the same amount of radioactive $^{26}$Al which has a 
much longer lifetime of $1.05\times10^6$ yr and whose decay produces a
$\gamma$-ray photon at 1.8 MeV.  Indeed, the COMPTEL instrument has detected
significant 1.8 MeV emission from the Vela region, whose origin has
previously been attributed to the Vela SNR (\cite{rf:drea95}) but later
become less certain (\cite{rf:drea99}).  However, the better coincidence of
the centroid of the COMPTEL 1.8 MeV feature with the location of GRO/RX J0852
prompts us to investigate if it is GRO/RX J0852 that is responsible for most
of the observed $^{26}$Al emission.

\section{The Nature of the Progenitor}

\begin{figure}
\psfig{figure=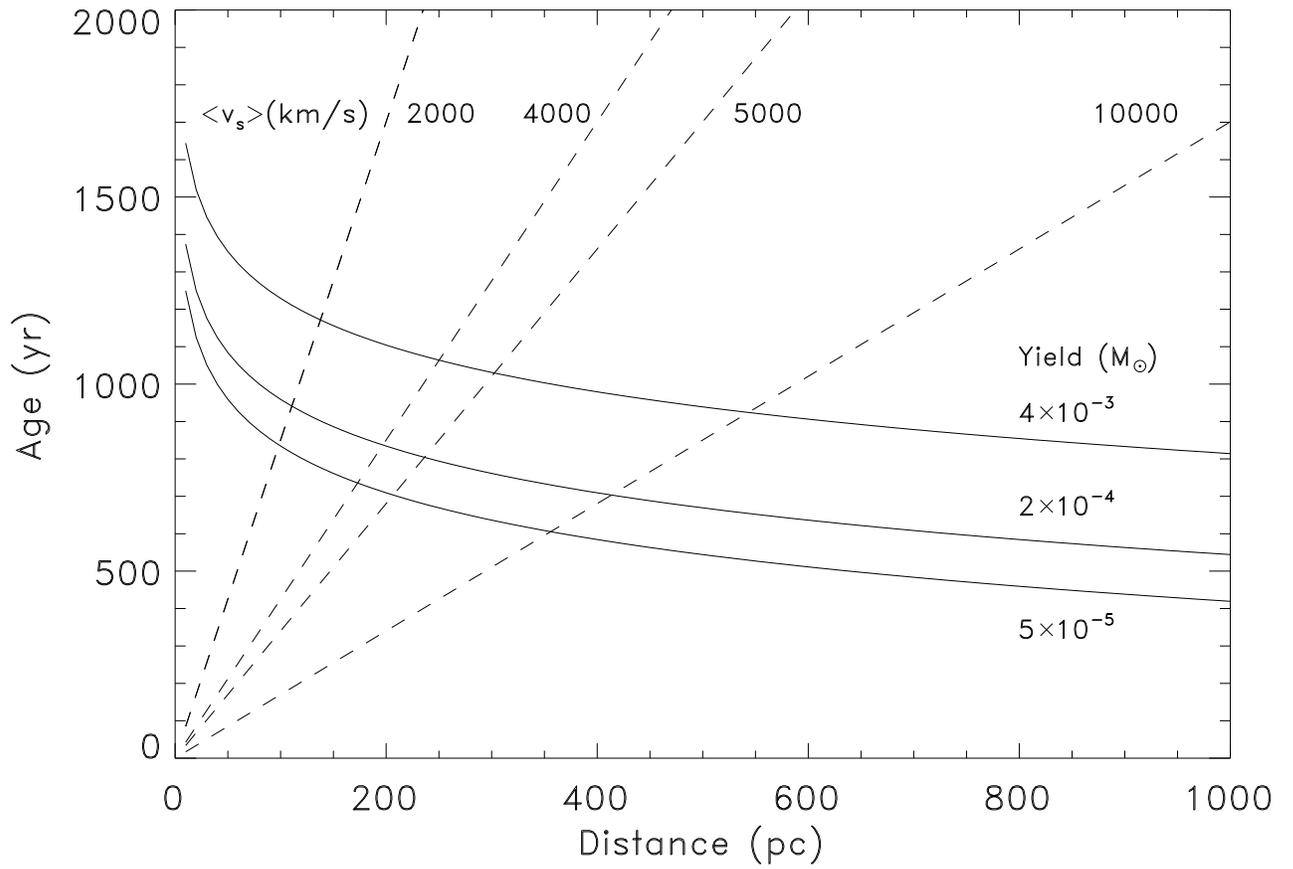}
\caption{ The constraint on the SNR distance and age by average shock
velocity and $^{44}$Ti yield from a single SNE.  \label{fg:age-dist} }
\end{figure}

The observed $^{44}$Ti line flux is $F = (3.8\pm0.7)\times10^{-5}$ cm$^{-2}$ 
s$^{-1}$ ($5.6\sigma$).  For a nominal $^{44}$Ti yield of $Y = 5\times10^{-5} 
M_\odot$ per supernova explosion (SNE, \cite{rf:ww94}; \cite{rf:ww95}; 
\cite{rf:tnh96}), the observed 1.16 MeV flux restricts the age of the source 
to be, 
\begin{equation}
t \le \tau\ln\left(\frac{Y}{4\pi d^2\tau m_{44}F}\right) = 840\pm20 \,\, 
 {\rm yr} 
\label{eq:t_obs}
\end{equation} 
at a distance of 100 pc.  In theoretical models, the yield ranges from a low 
of $5\times10^{-5}M_\odot$ to a high of $4\times10^{-3}M_\odot$. We calculate 
the constraint on the age versus distance for the range of the theoretical 
yields, as shown by the solid curves in Fig.~\ref{fg:age-dist}.

While the thermal nature of the bright northern rim is debatable, the thermal
X-ray emission from most of the shell is undoubtedly generated from shocks as
the supernova blast wave runs into the ambient medium.  The observed shell
temperature of 2.5 keV implies that the shock front is currently proceeding
at a relatively low velocity of 
\begin{equation} 
v_s = \left(\frac{16 kT}{3m_H}\right)^{1/2} = 1300^{+1020}_{-160} 
 \left(\frac{kT}{2.5^{+4.5}_{-0.7}\,{\rm keV}}\right)^{1/2} \,\,
 {\rm km\,\,s}^{-1}.  
\label{eq:v_s} 
\end{equation} 
The validity of such a direct conversion from the electron temperature to the
shock front speed requires the electrons to be in equilibrium with ions in
the post-shock region.  For collisionless shocks (as is the case for most
young SNRs), this condition may not always hold, if, for example, the
magnetic coupling between the ions and electrons is not effective enough.  In
such cases, the ion temperature, which reflects the true shock speed, could
be greater than the electron temperature (\cite{rf:sjd95}).  Therefore, $v_s$ 
calculated above is probably a lower limit; the real value could be greater 
by as much as a factor of 2-3.

Over the lifetime of the SNR, the mean expansion velocity $<v_s>$ may be
higher than $v_s$, depending on its progenitor type.  From arguments we
present below, $<v_s>$ is probably in the range of 2000--5000 km s$^{-1}$.  
For given $<v_s>$, the observed shell radius gives another set of constraints 
on the relationship between age and distance, as shown by the dashed lines in 
Figure~\ref{fg:age-dist}.  We see that the most probable value for the 
distance to the new SNR is in the range between 100 and 300 pc, and for the 
age is between 600 and 1100 years.  At a distance of 250 pc, the SNR has a 
radius of about 4.4 pc.

Now we discuss what possible progenitors are allowed by the observed and
derived properties of the SNR.  We first assume that the environment in which
the progenitor exploded has a uniform density $n_H= 0.05$ cm$^{-3}$,
consistent with the limit derived from the ROSAT data.  For an SNE in such a
medium, there is a universal solution for the early stages of SNR evolution
(\cite{rf:mt95}).  Immediately after the explosion, the blast wave expands
freely during the so-called ejecta-dominated phase.  The ejecta maintains a
constant expansion velocity with a maximum value of 
\begin{equation} 
v_e = \left(\frac{10E}{3M_e}\right)^{1/2} = 1.3\times10^4\; E_{51}^{1/2}\;
 M_{e,\odot}^{-1/2} \,\,{\rm km\, s}^{-1}, 
\label{eq:v_e} 
\end{equation} 
where $M_{e,\odot}$ is the mass of the ejecta in units of solar masses and
$E_{51}$ is the kinetic energy of the SNE in units of $10^{51}$ ergs.  For a
Type Ia supernova (SNIa) of $M_{e,\odot} \sim 1.4$, $v_e$ is more than 10,000
km s$^{-1}$.  For a Type II supernova (SNII) of $M_{e,\odot} \ge 10$, $v_e$
can be much smaller.  Therefore, if the progenitor of the new SNR is an SNIa,
the expansion must have been significantly slowed down which requires an old
age.  If the progenitor star is more massive than $30M_\odot$, the
free-expansion velocity of an SNII is close to the observed shock front
velocity $v_s$.

After the blast wave sweeps up about the same amount of mass as the ejecta,
the SNR evolution enters the Sedov-Taylor phase, during which time the SNR
shell velocity decreases with time as $v \propto t^{-2/5}$.  The timescale
for the onset of the Sedov-Taylor phase is (\cite{rf:mt95}), 
\begin{equation} 
t_{\rm ST} \sim 570 \; E_{51}^{-1/2}\; M_{e,\odot}^{5/6}\;
 n_{H,0.05}^{-1/3} \,\,{\rm yr.}  
\label{eq:t_ST} 
\end{equation} 
where $n_{H,0.05} = n_H / (0.05$ cm$^{-3}$).  Therefore, an SNII spends much 
more time in the free-expansion phase than an SNIa does.  Also, a higher 
ambient density of $n_H \sim 1$ cm$^{-3}$ would reduce $t_{\rm ST}$ by a 
factor of 2.7.

It immediately follows from eq.~(\ref{eq:v_e}) that if the progenitor of
GRO/RX J0852 was an SNIa, the SNR would have to be in the Sedov-Taylor phase
by now because of the observed small shock velocity.  However, for a typical
SNIa explosion with $\sim1\times10^{51}$ ergs of kinetic energy, it would
take about 
\begin{equation} 
t_{\rm Ia} \sim t_{\rm ST} \left(\frac{v}{v_e}\right)^{-5/2} = 
 1.5\times10^5\; E_{51}^{3/4}\; M_{e,\odot,1.4}^{-5/12}\; n_{H,0.05}^{-1/3}\; 
 v_{1300}^{-5/2} \;\,{\rm yr}
\label{eq:t_Ia} 
\end{equation} 
for the ejecta to slow down from an initial expansion velocity of 11,000 km
s$^{-1}$ to the observed 1300 km s$^{-1}$.  This age can drop to $10^4$ yr
if the real $v_s$ is a factor of 3 higher than that derived from the observed
electron temperature by eq.~(\ref{eq:v_s}), but it is still more than an oder
of magnitude longer than the data suggest.  For an SNIa model to work, the
only other alternative is then to increase the ambient density by a factor of
at least 10,000 to $\ge 500$ cm$^{-3}$.

\begin{figure}
\psfig{figure=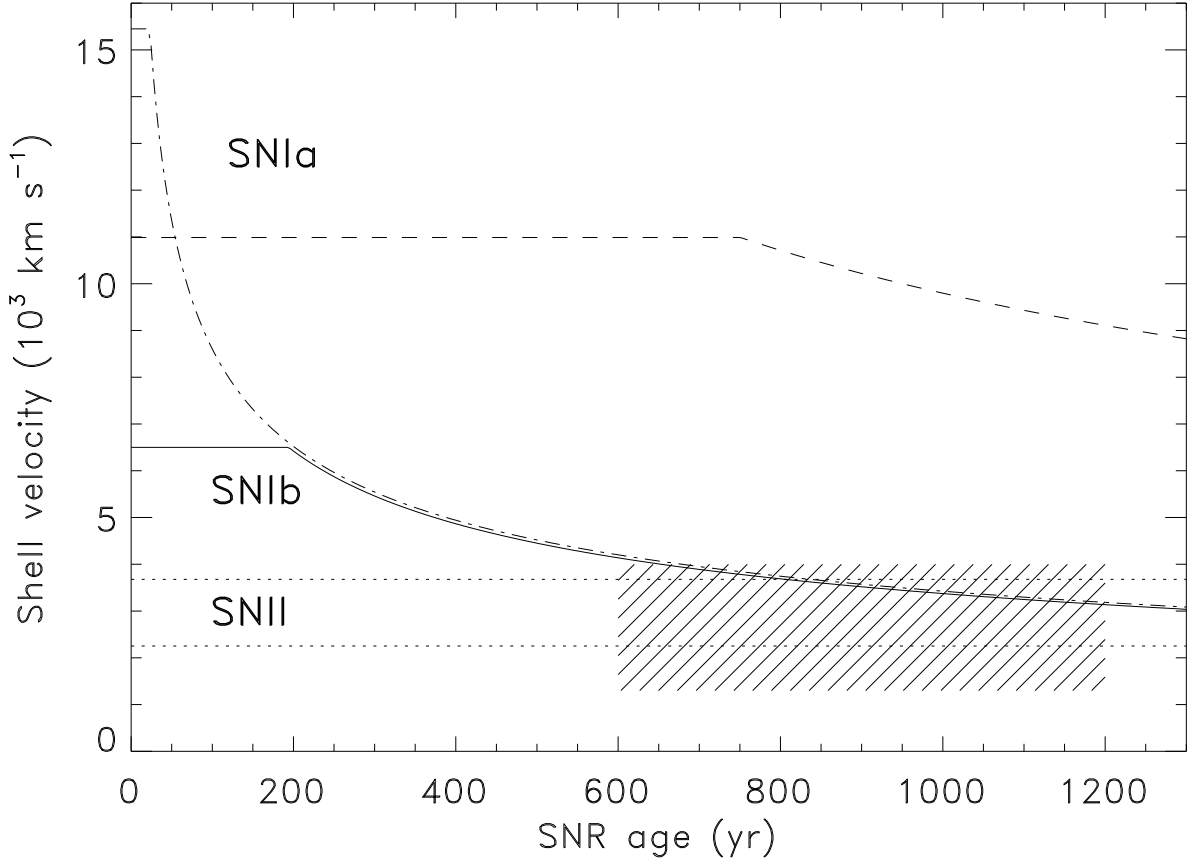}
\caption{ The theoretical SNR shell expansion velocity as a function of
the SNR age for SNIa, SNIb (including SNIc), and SNII models.  The input
parameters for the SNIa models are $E_{51} =1.0$, $M_{e,\odot} =1.4$, and
$n_H =0.05$ ({\it dashed} line) and $E_{51} =1.6$, $M_{e,\odot} =1.1$, and
$n_H =500$ ({\it dot-dashed} line).  Those for the SNIb model are
$E_{51}=2.5$, $M_{e,\odot} =10$, and $n_H =100$ ({\it solid} line), and for
the SNII models ($n_H = 0.05$) are $E_{51} =1.2$, $M_{e,\odot} =15$ ({\it
upper dotted} line), and $E_{51} =1.2$, $M_{e,\odot} =40$ ({\it lower dotted}
line).  The shaded area represents the region that is allowed by the
observations.  \label{fg:vel-age} }
\end{figure}

The above argument can be more clearly illustrated by considering the
theoretical SNR expansion velocity as a function of the SNR age shown in
Fig.~\ref{fg:vel-age}.  The constant velocity portion of the curves
represents the free-expansion (or ejecta-dominated) phase, followed by the
$v\propto t^{-2/5}$ Sedov-Taylor phase.  We have adopted the kinetic energy
for an SNIa to be around $1\times10^{51}$ ergs and for an SNII to be in the
range $1-2 \times 10^{51}$ ergs (\cite{rf:ww94}, 1995; hereafter WW94, WW95).
The ejecta mass of an SNIa varies from the standard model of $1.4 M_\odot$ to
special He-detonated models of $0.8M_\odot$ (WW94).  The ejecta mass of an
SNII is in the range between $8M_\odot$ and $40 M_\odot$.  The shaded area in
Fig.~\ref{fg:vel-age} represents the parameter space that is preferred by the
observations.  We have allowed a factor of 3 uncertainty in the current shell
expansion speed.

If we take the default ambient gas density as 0.05 cm$^{-3}$, we see from
Fig.~\ref{fg:vel-age} that the SNIa evolution track is totally inconsistent
with the observations.  To match the current expansion speed, the ambient
density has to be higher than 500 cm$^{-3}$ for the SNIa model to be
marginally acceptable.  In this case, we chose a He-detonation model from
WW94 with an ejecta mass of $1.1 M_\odot$, which has a $^{44}$Ti yield of
$\sim 2 \times 10^{-4} M_\odot$.  The mean expansion velocity for this case
is about 5100 km s$^{-1}$ at an age of 800 yr.  From Fig.~\ref{fg:age-dist}
we see that this model with very high density does have a solution with a
distance $\sim 240$ pc.

For the low ambient density condition, on the other hand, we see from
Fig.~\ref{fg:vel-age} that (1) there is a range of SNII models allowed, and
(2) all these models are still in the free-expansion phase, so $<v_s> = v_e$.
The model with the highest allowable free-expansion speed of $\sim 3700$ km
s$^{-1}$ ({\it upper dotted line}, Fig.~\ref{fg:vel-age}) has $E_{51} = 1.2$ 
and $M_{e,\odot} = 15$; it produces $5.7\times 10^{-5} M_\odot$ of $^{44}$Ti 
(WW95).  Fig.~\ref{fg:age-dist} then tells us that the source must be at a 
distance of about 150 pc and an age of about 700 yr. Other models with 
smaller expansion speed have much greater progenitor mass of 25 to 40 
$M_\odot$.  The lower dotted line of the SNII range in Fig.~\ref{fg:vel-age} 
corresponds to $E_{51} = 1.2$ and $M_{e,\odot} = 40$.  However, this model 
produces only $2.7 \times 10^{-11} M_\odot$ of $^{44}$Ti (WW95); in fact, the 
$^{44}$Ti yield of all the models in WW95 which have $v_e \le 3000$ km 
s$^{-1}$ is less than $10^{-9} M_\odot$.

Therefore, if the current SNR expansion speed is less than 2500 km s$^{-1}$,
the remnant of a viable core-collapse supernova model has also to be in the
Sedov-Taylor phase which requires a large ambient density of $n_H \ge 100$
cm$^{-3}$.  For reasons we discuss below, this condition dictates the source
distance to be at least 250 pc.  From Fig.~\ref{fg:age-dist} we see that at
such a distance the SNR mean expansion speed has to be $\ge 5000$ km
s$^{-1}$ because the $^{44}$Ti yield of these models is $<3 \times 10^{-4}
M_\odot$ (WW95).  Therefore, we need small ejecta mass and large kinetic
energy.  The lowest ejecta ($9.7M_\odot$) model of WW95 has a $^{44}$Ti yield
of $6\times10^{-5} M_\odot$ and also a relatively low kinetic energy, $1.3
\times10^{51}$ ergs.  It will produce a mean expansion speed of only 3700 km
s$^{-1}$ for $n_H =$ 300 cm$^{-3}$.  A more energetic explosion usually
requires a much more massive progenitor and thus even lower mean expansion
speed.  It is possible, however, that a massive star may undergo a supernova
explosion (resulting in a Type Ib or Type Ic supernova, for example) after it 
loses a significant fraction of its envelop because of strong stellar winds.  
While WW95's calculations did not take into account of this effect, we assume 
that it is conceivable that some core-collapse supernovae could have a 
kinetic energy as much as $2.5\times10^{51}$ ergs and at the same time an 
ejecta mass of $10M_\odot$.  If so, Fig.~\ref{fg:vel-age} shows that such a 
model (marked as SNIb) can produce a marginally acceptable evolution track in 
an ambient density of 100 cm$^{-3}$.  The mean expansion speed in this case 
is 5100 km s$^{-1}$, consistent with the required distance of 250 pc
(Fig.~\ref{fg:age-dist}).

\section{Discussions}

We have seen that whether the progenitor of GRO/RX J0852 is an SNIa or an
SNII, if the current SNR expansion speed, $v_s$, is less than 2500 km
s$^{-1}$, it is almost exclusively required that the SNE occurred in a high
density region, $n_H \ge 100$ cm$^{-3}$ for an SNII and $\ge 500$ cm$^{-3}$
for an SNIa.  The distance to the source in either case is about 250 pc and
most of the space along the line of sight still has a very low density of
$<<1$ cm$^{-3}$.  We now discuss whether the required high density medium
exists at the required direction and distance.

Along the general direction of GRO/RX J0852, the only possible candidate of a
high density region within 500 pc is the Gum Nebula.  While most of the Gum
Nebula is filled with ionized, low density gas (\cite{rf:ss93};
\cite{rf:rrj76}), recent reanalysis of the historical HI 21 cm data revealed
two distinct structures associated with the Gum Nebula (\cite{rf:rd97}).  One
is a large, $1.4\times10^5M_\odot$, disk of neutral gas which is 150 pc in
radius and $\sim50$ pc in thickness.  The other is a smaller, thick HI shell
of $2.1\times 10^4M_\odot$ with a radius of $\sim60$ pc (\cite{rf:dgcr92}),
roughly at the center of the disk.  Both the disk and the shell share a
distance of $500\pm100$ pc.  Therefore, the minimum distance to the edge of
the disk is about 250 pc.  The mean number density of the disk and the shell
is 1.6 cm$^{-3}$ and 0.95 cm$^{-3}$, respectively, far below the required
value for the model progenitor of GRO/RX J0852.  Thus, both the SNIa and SNIb
models require the presence of additional, much denser gas components at the
edge of the HI disk.  Because the Gum nebula was probably created by some
ancient supernova events, it is conceivable that the highest density regions
are located near the edge of the structure.

A recent CO survey has shown complex molecular cloud structure in the Vela
region (\cite{rf:mm91}).  The map shows that GRO/RX J0852 falls on the edge
of one of the CO hot spots but there is no morphological evidence that the
cloud is disturbed by the SNR.  The distances to the components of this
so-called Vela molecular ridge are estimated to be 1-2 kpc based on the
radial velocity profile, further suggesting that they may be unrelated to the
SNR.  If for any reason that these distance estimates were inaccurate and the
CO hot spot near GRO/RX J0852 had a distance of 250 pc instead of 1-2 kpc,
the maximum mean number density of this component would be even smaller than
the observed 16 cm$^{-3}$ (\cite{rf:mm91}), and fall short of the required
minimum density of 100 cm$^{-3}$.  Therefore, an extreme density of 500
cm$^{-3}$, along with it the SNIa origin of the new SNR, can be ruled out by
the HI and CO data; even the SNIb progenitor also becomes a difficult
proposition.

If GRO/RX J0852 was created by a SNII event about 800 yr ago, an unavoidable
consequence is that the same explosion which produced the observed $^{44}$Ti
will also synthesize certain amount of $^{26}$Al.  Specifically, for the SNIb
event favored by a small $v_s$ the $^{26}$Al yield is $3.6 \times 10^{-4}
M_\odot$ with a 1.8 MeV line flux of $\sim 6.7 \times 10^{-5}$ photons
s$^{-1}$ cm$^{-2}$ at a distance of 250 pc.  This flux is about a factor of 3
greater than the observed flux observed from the Vela region by COMPTEL
(\cite{rf:drea95}).  The $15 M_\odot$ SNII model favored by the condition
$v_s> 3500$ km s$^{-1}$, on the other hand, has an $^{26}$Al yield of
$4.3\times 10^{-5} M_\odot$, which would produce a 1.8 MeV line flux of
$2.2\times10^{-5}$ photons s$^{-1}$ cm$^{-2}$, comparable to the observed
value.  It is thus intriguing to postulate that GRO/RX J0852 is also a major
contributor to the Vela $^{26}$Al feature.  Indeed, recent reexamination of
the Vela data has argued against previous attribution of the Vela $^{26}$Al
flux to the Vela SNR (e.g., \cite{rf:drea99}).  The centroid of the Vela
$^{26}$Al feature, at $l= 267^\circ$ and $b= -1^\circ$, is $\sim4^\circ$ away
from the center of the Vela SNR but falls right on top of GRO/RX J0852.  The
combination of the positional coincidence and the good agreement between the
expected 1.8 MeV flux and the observed value strongly favors this
interpretation.  Since the 1.8 MeV Vela feature is probably slightly
extended, it seems that both the Vela SNR, or a local concentration of the
massive-star formation tracers (\cite{rf:drea99}), and GRO/RX J0852 have
contributed while the majority of the flux comes from the latter.

\section{Conclusions}

We have investigated the parameter space allowed by the observed properties
of the newly discovered SNR GRO/RX J0852, assuming that the COMPTEL $^{44}$Ti
source and the ROSAT SNR are the same source.  We conclude that the likely
progenitor of this new SNR is a massive star and its precise type depends on
the correct determination of the current SNR shell expansion speed $v_s$.  It
is most probably that the true value of $v_s$ is more than a factor of 2
greater than that derived from the shell electron temperature using X-ray
data and the progenitor is a $15M_\odot$ SNII with a moderate SNE kinetic
energy.  The age of the source is $\sim 700$ yr and the distance is $\sim
150$ pc.  If the current $v_s$ were small, on the other hand, a much more
energetic SNIb or SNIc event would be required to have exploded 800-900 yr
ago in a high density medium about 250 pc away.  This latter interpretation,
however, has the difficulty of finding observationally the required high
density gas at the required distance.  The recently discovered cold neutral
gas structure associated with the nearby Gum Nebula seems not to be able to
provide the required high density.

It is likely that the nearest SNE less than 1000 yr ago in the Vela region
produced both the $^{44}$Ti and $^{26}$Al emission features detected by
COMPTEL.  This offers an exciting new avenue to test the core-collapse SNE
models since the explosive yields of $^{44}$Ti and $^{26}$Al are related.
Because of its much shorter lifetime, $^{44}$Ti is usually better suited for
discovering young SNRs.  However, since it just so happened that the age of
GRO/RX J0852 is about 8-10 times the lifetime of $^{44}$Ti, its $^{44}$Ti
flux has dropped by a factor of 3000--10,000 from its peak value and now
almost equals that of $^{26}$Al which has a decay lifetime that is $\sim
10,000$ times longer.  It is thus worth noting that for SNRs older than about
1200 yr, $^{26}$Al is a better radionuclide to use for SNR searches.

\acknowledgements We thank Roland Diehl for communicating with us about the
new analysis of the COMPTEL 1.8 MeV feature in Vela and are very grateful to 
the anonymous referee for his/her insightful comments.

\end{document}